\documentclass[twocolumn,pra,aps,superscriptaddress,showpacs]{revtex4-1}
\usepackage{amsmath}
\usepackage{amssymb}
\usepackage{amsfonts}
\usepackage{amssymb}
\usepackage[dvips]{graphicx}
\usepackage{subfigure}
\usepackage{dcolumn}
\usepackage{txfonts}
\usepackage{bm}
\usepackage{makeidx}
\usepackage{color}
\usepackage{mathtools}
\usepackage{threeparttable}
\usepackage[colorlinks,linkcolor=blue,anchorcolor=blue,citecolor=blue,urlcolor=blue]{hyperref}
\usepackage{stackrel}
\usepackage{esint}
\usepackage{mathtools}
\usepackage{wasysym}

\begin{document}

\title{Optical Levitation of Nanodiamonds by Doughnut Beams in Vacuum}
\author{Lei-Ming Zhou}
\address{Beijing Computational Science Research Center, Beijing 100193,
China}
\author{Ke-Wen Xiao}
\address{Beijing Computational Science Research Center, Beijing 100193,
China}
\author{Jun Chen}
\address{Institute of Theoretical Physics and Collaborative Innovation Center
of Extreme Optics, Shanxi University, Shanxi, China}
\author{Nan Zhao}
\email{nzhao@csrc.ac.cn}
\address{Beijing Computational Science Research Center, Beijing 100193,
China}

\date{\textcolor{blue}{\today }}

\begin{abstract}
Optically levitated nanodiamonds with nitrogen-vacancy centers
promise a high-quality hybrid spin-optomechanical system. However, the
trapped nanodiamond absorbs energy form laser beams and causes thermal damage in vacuum.
It is proposed here to solve the problem by trapping a composite particle (a
nanodiamond core coated with a less absorptive silica shell) at the center
of  strongly focused doughnut-shaped laser beams. Systematical study on the
trapping stability, heat absorption, and oscillation  frequency  concludes
that the azimuthally polarized Gaussian beam and the linearly polarized
Laguerre-Gaussian beam ${\rm LG}_{03}$ are the optimal choices. 
With our proposal, particles with strong absorption coefficients can be trapped without obvious heating and, thus, the spin-optomechanical system based on levitated nanodiamonds are made possible in high vacuum  with the present experimental techniques. 
\end{abstract}

\maketitle

\textit{Introduction-.}\label{sec:Introduction} 
By trapping, detecting and manipulating nano- and micro-particles \cite{GrierNature2003}, optical
tweezers are widely used in biophysics \cite%
{AshkinOL1986,AshkinNature1987,AshkinScience1987}, colloidal sciences \cite{DholakiaRMP2010Colloquium}, chemistry, microfluidic dynamics \cite{DholakiaCSR2008Optical}, and  fundamental physics \cite%
{ChangPNAS2010Cavity,IsartPRL2011large,
LiTCScience2010,LiTCNPhysics2011Millikelvin,KheifetsScience2014,GieselerPRL2012Subkelvin,GieselerNNano2014Dynamic,MillenNNano2014Nanoscale,JainPRL2016DirectRecoil}. %
Because of the wide applicability and high tunablity of the optically levitated systems, several schemes \cite{YinIJMPB2013optomechanics} were proposed to realize
the ground-state cooling \cite{RablPRB2009strong}, 
to search for non-Newtonian gravity \cite{GeraciPRL2010Short} and to detect
gravitational wave \cite{ArvanitakiPRL2013Detecting}. Particularly, it brings about more interesting phenomena and novel applications \cite{ScalaPRL2013matter,ZhaoPRA2014massspectrometer} when the trapped particles have internal degrees of
freedom (such as spins or electric dipoles) and enter the quantum regime.

Optically levitated nanodiamonds with nitrogen-vacancy (NV) centers \cite{YinPRA2013large,YinSCPMA2015Hybrid,Neukirch2015NPhotonicsMulti,HoangNCom2016,FrangeskouArXiv2016} are one
of the most promising candidates for implementing a spin-optomechanical hybrid system. In principle,
this system can have both long spin coherence time and high quality factor of mechanical
oscillation in vacuum. The electron spins of NV centers were shown to have
long spin coherence time (in the order of $10^{2}~\mathrm{\mu s}$) even in
nanodiamonds of diameter about $20~\mathrm{nm}$ \cite{KnowlesNMater2014observing}. When trapped
in high-vacuum, the dielectric particles are predicted to have ultra-high
quality factor $Q$ larger than $10^{10}$ \cite{GeraciPRL2010Short,YinIJMPB2013optomechanics,GieselerNPhys2013}. Researchers have trapped diamond particles and
observed the signal from NV centers in liquid \cite%
{HorowitzPNAS2012Electron,GeiselmannNNano2013}, in air \cite%
{NeukirchOL2013} and very recently in vacuum with pressure down to $\mathrm{\sim kPa}$ \cite%
{Neukirch2015NPhotonicsMulti,HoangNCom2016} and $\mathrm{\sim 100~Pa}$ \cite{FrangeskouArXiv2016}. 

Realizing high quality mechanical oscillation requires trapping the
particles in high vacuum (e.g, $\mathrm{10^{-6}~Pa}$) to get $Q~\sim~10^{10}$.
However, the high-vacuum condition usually causes
the thermal damage problem, and experimentally
trapping a nanodiamond in high vacuum is still very challenging. Nanodiamonds will absorb energy from the trapping laser beams
due to the intrinsic defects \cite{FrangeskouArXiv2016} and the inevitable
imperfections or graphitization \cite{RahmanSR2016} on diamond surface. The
absorbed energy can hardly be dissipated in a high-vacuum environment, and
the nanodiamonds will be quickly heated up significantly \cite%
{Neukirch2015NPhotonicsMulti,HoangNCom2016,FrangeskouArXiv2016}, which is unfavorable to the
defect centers, or even burns out the diamond particles. Improving the
purity of trapped nanodiamonds is one way to reduce the heat
absorption \cite{FrangeskouArXiv2016}. Here, we study
an alternative way by engineering the trapping beams, which
is in principle applicable to much wider range of particles.

\begin{figure}[bp]
\includegraphics[width=8.5cm]{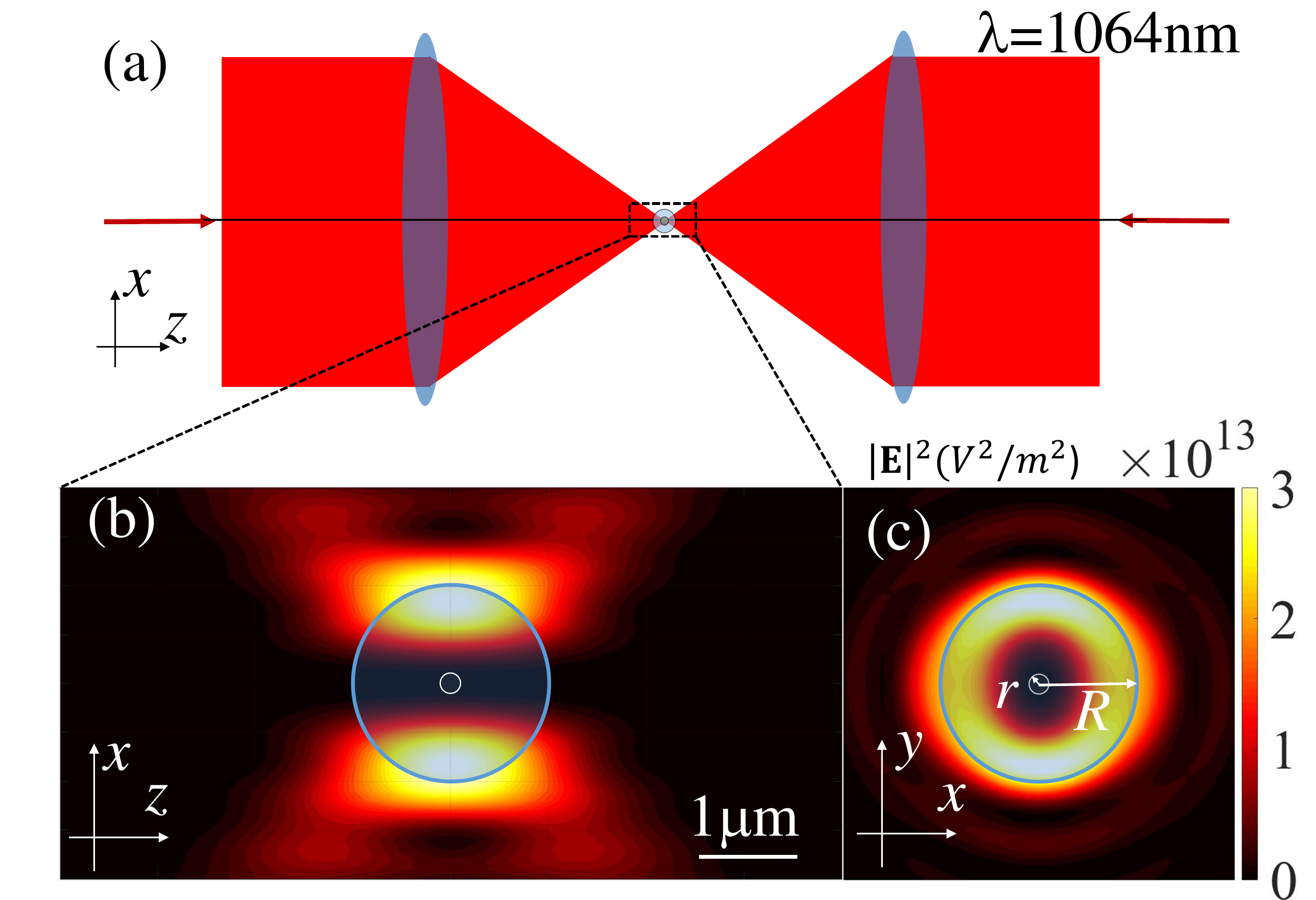}
\caption{ (a) The schematic illustration of the system. A nanodiamond coated
with a silica shell is levitated in an optical trap formed by two incoherent
strongly-focused counter-propagating beams. (b) (c) Front view and side view
of the intensity distribution of the two focused linearly polarized $\mathrm{%
LG}_{03}$ incident beams in the focal region. The circles indicate the
composite particle with the core radius $r=100~\mathrm{nm}$ and the shell
radius $R=1~\mathrm{\protect\mu m}$. Two incident beams are with total power $%
100~\mathrm{mW}$.}
\label{fig:trap-structure}
\end{figure}

It is proposed here to solve the thermal damage problem by trapping a silica-coated
nanodiamond with doughnut beams [e.g., the Laguerre-Gaussian (LG) beams, see
Fig.~\ref{fig:trap-structure}]. Our proposal is based on the
following two observations. Firstly, recent experiments \cite%
{LiTCScience2010} show that micro-sphere made of silica can be trapped in
high-vacuum without strong heat absorption because of the low absorption
coefficient. Secondly, it is well-known that the cross-section intensity
distribution of the doughnut beam has a dark region at the beam center.
Nanodiamonds can be coated with a silica shell \cite%
{HaartmanJMCB2013,Neukirch2015NPhotonicsMulti}, forming a core-shell
structure. When the dark region of the beam coincides with the diamond core,
the heat absorption will be significantly suppressed. Trapping and manipulating particles with doughnut beams have been investigated \cite{HeJMO1995,GahaganOL1996,HeckenbergOV1999mechanical,GanicOE2005,OtsuOR2015,CaoOE2016spin,HeckenbergOV1999mechanical} in liquids. Here, we focus on the problem in vacuum which has low friction and low dissipation.

\begin{figure}[tbp]
\includegraphics[width=8.4cm]{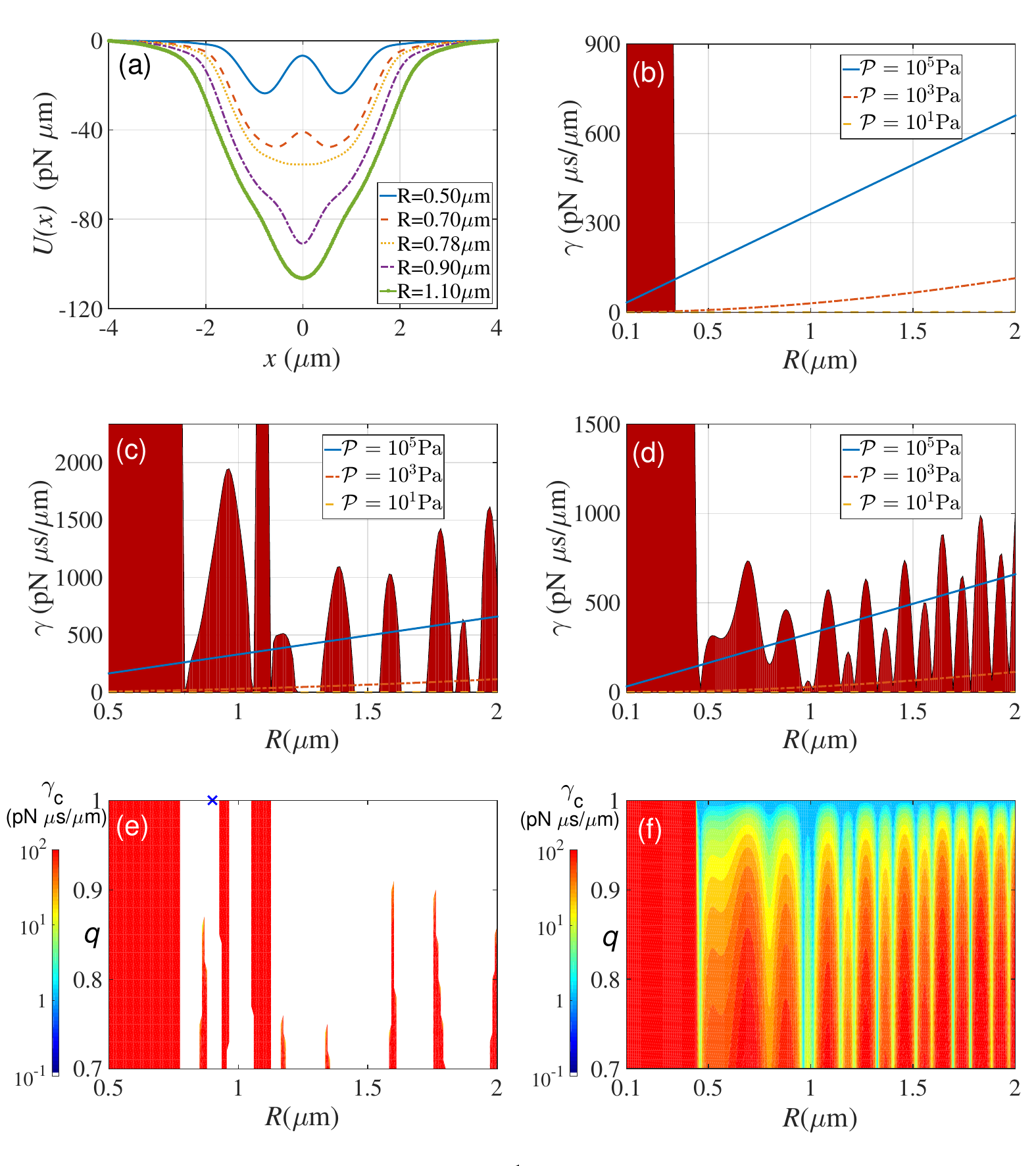}
\caption{ (a) The transverse optical potential $U(x)$ for a core-shell
spherical particle in a single linearly polarized $\mathrm{LG}_{03}$ beam
with power $P=100\ \mathrm{mW}$. (b)-(d) The transverse stability diagram of a
core-shell particle trapped in an azimuthally polarized beam, a
linearly polarized $\mathrm{LG_{03}}$ beam and a right circularly polarized
LG beam, in turn. The white regions are stable. The Stokes friction coefficients $\protect\gamma$ under different
pressure $\mathcal{P}$ are also plotted with lines for reference. (e)
The minimum friction coefficients ${\protect\gamma}_c$ required to stabilize
the optical trap formed by the dual-beam optical tweezers with the $\mathrm{%
LG_{03}}$ beams, as functions of the power mismatch ratio $q$ and particle
size $R$. The blue `$\times$' denotes a stably trapped sphere for light absorption suppression and equilibrium temperature simulation later. (f) The same as (e), but for the right circularly polarized LG
beams. Other parameters used here are $\protect\lambda=1064\ \mathrm{nm},$ $%
\mathrm{NA}=0.95$ and each beam with power $P=100\ \mathrm{mW}$ for all
figures. }
\label{fig:PhaseDiagram}
\end{figure}

The key problem is under what condition that the composite particle can be
stably trapped in vacuum with the core lying in the dark region. We
numerically solve the electromagnetic wave scattering problem and show that
the gradient force from a doughnut beam can form a single-well potential, as
long as the particle size exceeds a critical radius $R_{\mathrm{trans}}$
[see Fig.~\ref{fig:PhaseDiagram}(a)]. Furthermore, the doughnut beam usually
carries non-zero orbital angular momentum (OAM) and has different
polarizations. The OAM and polarization affect the trapping
stability and heat absorption of the particle, particularly in the case of
the strongly focused beam. We systematically investigate trapping effect of
doughnut beams with various OAM (e.g., the $\mathrm{LG}_{0l}$ beams) and
polarization (e.g., linearly or circularly polarized LG beams and the
cylindrical vector beams \cite{JonesBook2015optical}).
By comparing the trapping stability, the heat absorption, and the
oscillation frequency of the composite particle trapped in different types
of doughnut beams, we conclude that the azimuthally polarized Gaussian beam
and the linearly polarized $\mathrm{LG}_{03}$ beam are the optimal choices
for implementing the hybrid spin-optomechanical system in vacuum.

\textit{Trapping stability-.} \label{sec:Trapping stability} We consider a
dual-beam optical tweezers system as shown in Fig.~\ref{fig:trap-structure}%
(a). Two incoherent counter-propagating laser beams of wavelength $\lambda$
are focused by two identical lenses with numerical aperture
NA. When the two beams with same parameters except the
directions are well-aligned, the $z$ direction scattering forces from the
two beams cancel each other, and the gradient forces form an optical trap
near the focal point in three dimensional space.

We start the discussion from considering the strongly focused $\mathrm{LG}_{0l}
$ beams ($l>0$). The focused beams violate the paraxial condition, and we
perform numerical calculations of the focal field, following the theory
developed by Richards and Wolf \cite{Novotny2012principles,NgPRL2010}.
Figures~\ref{fig:trap-structure}(b) and \ref{fig:trap-structure}(c) show the
light intensity distribution in the focal region. Similar to the paraxial
case, there is a dark region along the beam propagating axis (the $z$ axis).
However, in contrast to the paraxial beams whose OAM are usually
well-defined and separated with the polarization degree of freedom, the
orbital and polarization degrees of freedom are highly mixed around the
focal point. More importantly, we will show that the trapping stability and
the heat absorption are sensitive to the choice of the OAM and the
polarization of the focused beams.

The focused beams provide an optical trap to the particle. As an example,
Fig.~\ref{fig:PhaseDiagram}(a) shows the trapping potential $%
U(x)=-\int_{-\infty}^{x}F_{x}(x^{\prime})dx^{\prime}$ along the $x$
direction for incident $\mathrm{LG}_{03}$ beams, where $F_{x}$ is the $x$
component of the optical force $\mathbf{F}$ when the particle is displaced
along the $x$ axis {\footnote{noted that this is the phenomenological
potential which is defined without distinguishing the gradient and
scattering force}}. Generally speaking, small particles tend to be trapped
at the position of maximal intensity. For incident LG beams, a small
particle (e.g., with radius $R\ll \lambda$) will be confined in the region
of the bright ring, corresponding to a double-well potential. With
increasing particle size, the trapping potential is gradually changed to a
single-well. The transition radius $R_{\mathrm{trans}}$ from double-well to
single-well is comparable to the radius of the bright ring. A composite particle with radius $R>R_{\mathrm{trans}}$ will
be trapped around the equilibrium position with its core locates in the dark
region of the focused LG beams.

The LG beams carry OAM, which accelerates the particle in the azimuthal
direction and strongly affects the trapping stability \cite{NgPRL2010}. The
trapping stability is determined by the force constant matrix $\mathbf{\mathbb{K}}%
=\nabla\mathbf{F}$ of the focused beam at the equilibrium position (when $%
R>R_{\mathrm{trans}}$). Because of the OAM of the LG beams, the $y$%
-component of the radiation force $F_{y}$ is nonzero, when the particle
displacement is along the $x$ direction. In this case, the trap stiffness $%
K_{i}$ (i.e., the eigen values of the force constant matrix $\mathbf{\mathbb{K}}$ for
$i=1,2$ or $3$) can be complex numbers \cite{NgPRL2010}. When ${\rm Re}[K_i]<0$, the trap is single potential well in the focal point.
However, if $\mathrm{Im}[K_{i}]\neq0$, the trap is unstable unless the
environmental damping is larger than a critical value $\gamma_{\mathrm{c}}$.
Figure~\ref{fig:PhaseDiagram}(c) shows a typical stability diagram of the
transverse motion of a core-shell particle trapped in a single $\mathrm{LG}%
_{03}$ beam. For the applications of optical tweezers in high vacuum, we are
interested in the absolutely stable regions (ASRs), where the particle can
be trapped in the absence of any damping ($\gamma=0$). In the case of the
linearly polarized $\mathrm{LG}_{03}$ beam, there are several ASRs (e.g.,
around $R=1.5~\mathrm{\mu m}$),

\begin{figure}[btp]
\includegraphics[width=8.4cm]{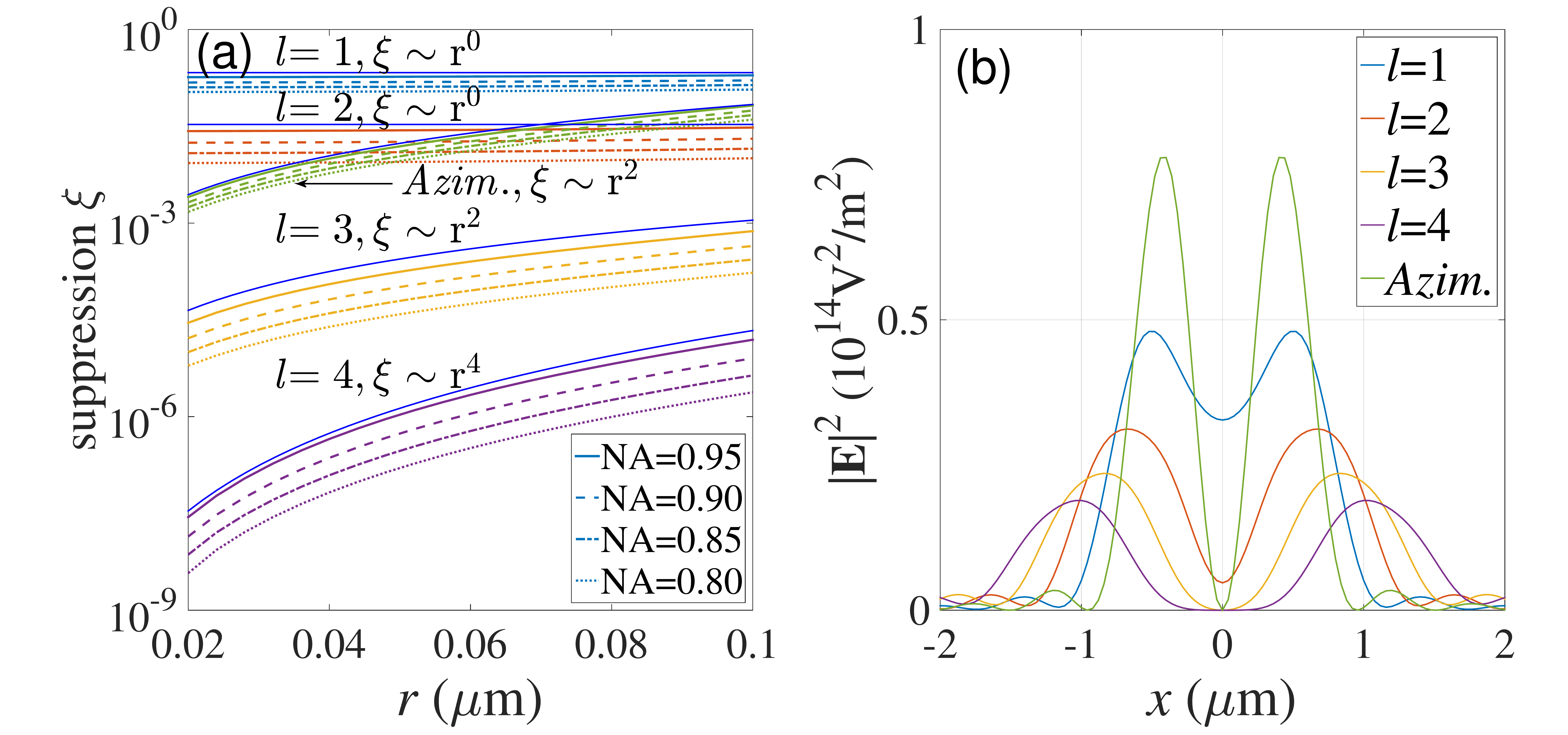}
\caption{ (a) Light absorption suppression ratio $\protect\xi$ for a
core-shell sphere at the center of azimuthally and linearly polarized beams (%
$\mathrm{LG}_{0l}$ beams) relative to Gaussian beams, with different core
radius $r$ and numerical aperture $\mathrm{NA}$. Blue solid lines are for our approximate analytical results. $R=900\ \mathrm{nm}$, refractive index $n_{\mathrm{diamond}}=2.418+0.001i$ and $n_{\mathrm{%
sillica}}=1.458$ (absorption eliminated in this figure). (b) Field $|\mathbf{%
E}|^{2}$ distribution of incident strongly focused azimuthally and linearly
polarized beams ($\mathrm{LG}_{0l}$ beams polarized along $x$ axis). All
beams are with power $P_1=P_2=50\ \mathrm{mW}$ focused by the lens of $%
\mathrm{NA}=0.95$.}
\label{fig:heat_absorption}
\end{figure}

The trapping stability depends on the beam polarization. Similar
calculations are made for the azimuthally and circularly polarized beams
[see Figs.~\ref{fig:PhaseDiagram}(b) and \ref{fig:PhaseDiagram}(d)]. Being
different from the linearly polarized beams, no ASR appears in the
circularly polarized case. A finite damping rate $\gamma$ due to the
environment (e.g., the collisions with the residual molecules) is necessary
to maintain a stable trapping. In contrast, the azimuthally polarized beam
provides a stable trap as long as the particle size exceeds the critical
radius.
This can be understood because the azimuthally polarized beam indeed does not carry OAM.

Counter-propagating beams with the same OAM improve the trapping stability.
When the particle is displaced along $x$ direction, the $x$ components
radiation force of the two beams add up, while the $y$ components are
canceled.
Figure~\ref{fig:PhaseDiagram}(e) shows the stability diagram for two
counter-propagating $\mathrm{LG}_{03}$ beams, with powers $P_{1}$ and $%
P_{2}=qP_{1}$ (with the power mismatch ratio $q$). The ASR is enlarged when
the ratio $q$ approaches to unity. For example, particles with radius $R$
around $0.9~\mathrm{\mu m}$ and $R>1.3~\mathrm{\mu m}$ can be stably trapped
even when the beam intensities are not perfectly matched (e.g., in the
region with $0.9<q\le1$). In the circularly polarized case, the
counter-propagating beams reduce the critical damping coefficient.
Unfortunately, the ASR appears only when the beam intensities are exactly
matched (i.e., $P_1=P_2$ or $q=1$). Accordingly, in terms of trapping
stability, we conclude that the azimuthally polarized Gaussian beam and the
linearly polarized LG beams are appropriate for the dual-beam optical
tweezers in high vacuum. The cases for doughnut beams with different OAM and polarization have been investigated systematically and more details can be seen in supplementary information.

\textit{Heat absorption-.}\label{sec:Light absorption suppression} Having
discussed the trapping stability, now we turn to the suppression of the heat
absorption. Figure~\ref{fig:heat_absorption}(a) shows the absorption
coefficients $c_{\mathrm{abs}}$ of the a composite particle in azimuthally
polarized Gaussian beam and linearly polarized $\mathrm{LG}_{0l}$ beam with
different OAM index $l$. All the absorption coefficients are normalized by
the absorption coefficient $c_{\mathrm{abs},0}$ of the same particle in a
Gaussian beam, and are thus defined as suppression ratio
\begin{equation}
\xi\equiv \frac{c_{\mathrm{abs}}}{c_{\mathrm{abs},0}}.
\end{equation}

The heat absorption depends on both polarization and OAM. In the azimuthally
polarized beam, the heat absorption of the particle is reduced by a factor
of $10^2$ to $10^{3}$, depending on the size of the absorptive core. While,
for the LG beams, the suppression ratio $\xi$ has qualitatively different
behavior for different OAMs. In the cases of $l=1$ and $l=2$, the
suppression effect is relatively weak and the ratio $\xi$ is independent on
the radius of the core. However, for $l\ge3$, the heat absorption is
significantly reduced ($\xi < 10^{-3}$ for core radius $r<100~\mathrm{nm}$),
and follows a power-law dependence of the core size.

\begin{figure}[tbp]
\includegraphics[width=8.4cm]{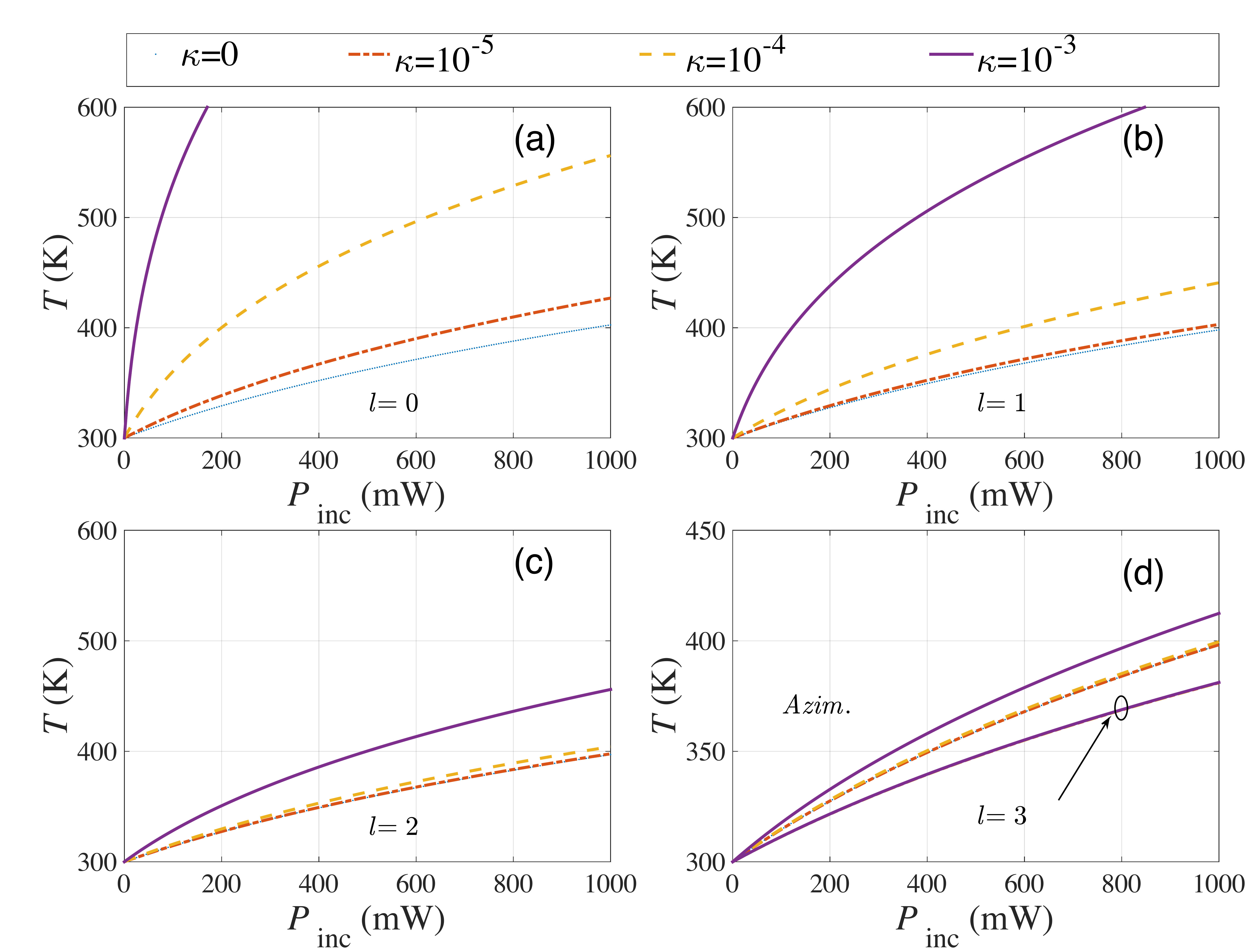}
\caption{The equilibrium temperature $T$ versus the incident laser power $P_{inc}$
for a core-shell sphere trapped in linearly polarized $\mathrm{LG}_{0l}$
beams and azimuthally polarized beam. Cases with (a) $l=0$, (b) $l=1$, (c) $%
l=2,$ (d) $l=3$ and (d) azimuthally polarized beam are shown. }
\label{fig:T_Pinc_k}
\end{figure}

Different behavior of the heat suppression ratio originates from the
intensity distribution of the incident strongly focused beams. Figure~\ref%
{fig:heat_absorption}(b) shows the intensity distribution along the $x$ axis
of the azimuthally polarized beam and LG beams with different OAMs.
For the typical core radius $r<100~\mathrm{nm}$, we expand the
intensity distribution into power series of $kx$ (in the $x$
direction for example, and $kx\ll1$ with the wave number $k=2\pi/\lambda$)
around the equilibrium position (i.e., the beam center). We
find that, for the azimuthally polarized beam, the intensity exactly
vanishes at the beam center ($x=0$) and increases quadratically as
increasing $x$, i.e., $I_{\mathrm{a}}\approx A_{\mathrm{a}}(kx)^2$. For the
strongly focused $\mathrm{LG}_{0l}$ modes, the intensity reads
\begin{equation}
I_l(x)\approx A_l (kx)^{n_l},\quad \text{for integer } l\ge1.
\end{equation}
Here, $A_{\mathrm{a}}$ and $A_{l}$ are the expansion coefficients. Being
different from the paraxial LG beams, the $\mathrm{LG}_{01}$ and the $%
\mathrm{LG}_{02}$ modes have finite intensity at the focal point (i.e., with
constant leading terms with $n_1=n_2=0$). For the $\mathrm{LG}_{0l}$ modes
with $l\ge3$, it is absolutely dark at the beam center [$I_{l\ge3}(0)=0$]
and the expansion power index depends on the OAM index $l$ as $n_l = 2(l-2)$%
. This explains the power-law behavior of the heat suppression ratio in Fig.~%
\ref{fig:heat_absorption}(a), and suggests that the azimuthally polarized
beam and the $\mathrm{LG}_{0l}$ modes with $l\ge3$ are good candidates for
solving the thermal damage problem. Details of the intensity expansion and suppression approximations can be seen in supplementary information. Systematically investigation for right circularly polarized beams has also been included.

With the absorption coefficient $c_{\mathrm{abs}}$, we estimate the
equilibrium temperature $T$ of the composite particle trapped by different
beams. We consider the vacuum environment of temperature $T_{0}$, where
radiation is the dominating heat transfer mechanism. The absorbed energy
from the beams is balanced by the black-body radiation as
\begin{equation}
c_{\mathrm{abs}}P_{\mathrm{inc}}+\sigma AT_{0}^{4}=\sigma AT^{4},
\label{eq:temperature}
\end{equation}
where $P_{\mathrm{inc}}$ is the total incident power of the
counter-propagating trapping beams, $\sigma$ is Stefan's constant and $%
A=4\pi R^{2}$ is the surface area of the sphere. Here, we have neglected the
heat dissipation due to the surface conduction $j=k_{\mathrm{s}} A(T-T_{0})$%
, where $k_{\mathrm{s}}$ is the surface conduct coefficient due to the
residual gas. Accordingly, Eq.~(\ref{eq:temperature}) gives an estimation of
the upper bound of the equilibrium temperature. Figure~\ref{fig:T_Pinc_k}
shows the equilibrium temperature $T$ of a stably trapped particle (denoted as a `$\times$' in Fig.~\ref{fig:PhaseDiagram}(e), $R=900 {\rm nm}$) as the function of incident power $P_{\mathrm{inc}}$. We assume
that the shell of the composite particle is made of silica with absorption
coefficient $100\ \mathrm{dB/km}$, corresponding to the imaginary part of
the refractive index $\kappa_{\mathrm{shell}}=2\times10^{-9}$. The diamond
core could be very absorptive due to the intrinsic defects and the imperfect
surface. We consider the imaginary part of refractive index of the diamond
core ranging from $\kappa_{\mathrm{core}}\sim 10^{-3}$ to $10^{-5}$. For the
fundamental Gaussian mode (i.e., $l=0$), the temperature increases
dramatically as increasing the incident power. However, when trapped by the
doughnut beams, particularly the azimuthally polarized beam and the $\mathrm{%
LG}_{03}$ mode [see Fig.~\ref{fig:T_Pinc_k}(d)], the diamond core has
negligible contribution to the temperature increasement. The composite particle
can afford much stronger power (up to the order of Watt) of the trapping
beams, without significant heating effect.

\begin{figure}[tbp]
\includegraphics[width=8.4cm]{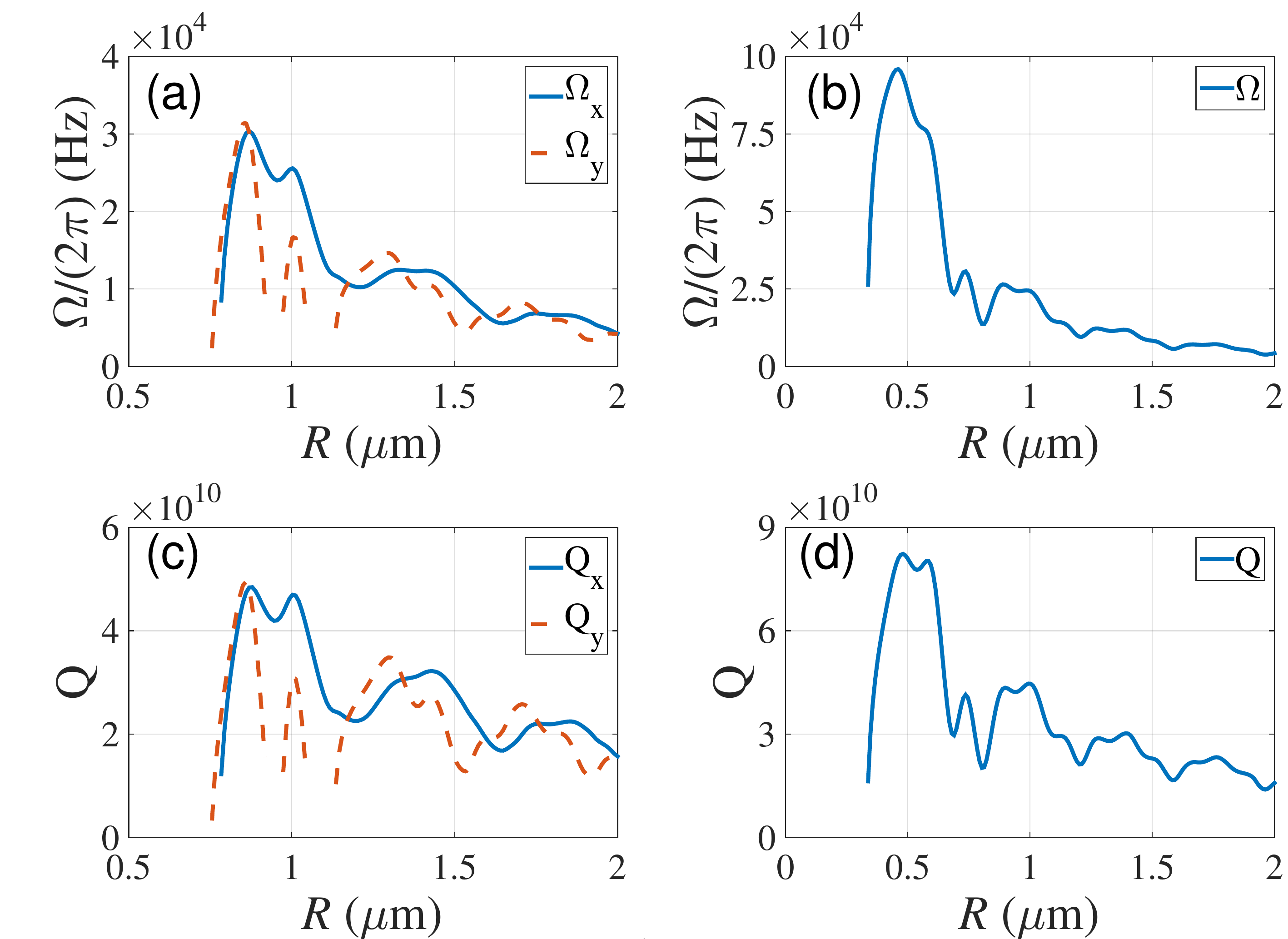}
\caption{The oscillation frequency $\Omega/(2\protect\pi)$ of the sphere in (a) linearly polarized $\mathrm{LG_{03}}$
beams and (b) azimuthally polarized beams.  And the mechanical
quality factor $Q$ of the sphere in (c) linearly polarized $\mathrm{LG_{03}}$
beams and (d) azimuthally polarized beams under pressure $10^{-6}\ \mathrm{Pa}$%
. All beams are with total power $100\ \mathrm{mW}$ and $\mathrm{NA}=0.95$.}
\label{fig:Freq_Qm}
\end{figure}

In this part, we discuss two figures of merit of the system, namely,
the trapping frequency $\Omega =\sqrt{K/M}$ and mechanical quality factor $%
Q=\Omega /\Gamma $ of the mechanical oscillation. Here $K$ is the force
constant, $M$ is the mass of oscillator, and $\Gamma =\gamma /M$ is the
damping coefficient with Stokes friction coefficient $\gamma $ due to the
residual gas. Figures~\ref{fig:Freq_Qm} presents
the frequency $\Omega $ and the quality factor $Q$ for the azimuthally
polarized beam and the linearly polarized $\mathrm{LG}_{03}$ beam, respectively. For a given
particle size, the azimuthally polarized beam provides a trapping frequency
in the order of $100~\mathrm{kHz}$, much higher than that of the $\mathrm{LG}_{03}$ mode. While the later creates an optical trap
with non-degenerate frequencies, in the order $10~\mathrm{kHz}$, in the $x$
and $y$ directions. The quality factor of the mechanical oscillation $Q$ is
inversely proportional to the damping coefficient $\gamma $ due to the
residual gas. The quality factor reaches $Q\sim 10^{10}$, with a residual
pressure $\mathcal{P}\mathrm{=10^{-6}}~\mathrm{Pa}$. Reducing the pressure
can further increase the quality factor, to the best figure of merit of this opto-mechanical system.

\textit{Conclusion}-.\label{sec:Conclusion} Optical tweezers achieved great
success in the past years in broad research fields ranging from biology,
statistical physics, to microchemistry. However, the heat absorption
problem prevents the exciting applications in the low-dissipation vacuum
environment. We propose to use doughnut beams and core-shell particles to
implement stable optical tweezers system in high vacuum. The low-absorptive
shell (e.g., the silica shell) plays a role of `sample-holder', which
interacts with the trapping beam and provides the radiation force for
levitation. While the `sample' could be more general particles (not
necessarily to be nanodiamonds). Once loaded in the sample-holder, no matter
how strong absorptive it is, the core particle has little chance to see the
trapping beam. With a systematic study of the physical effect of the beam
OAM and polarization on the trapping stability and heat absorption, we
provide a comprehensive solution to the heat absorption problem and make
optical tweezers a powerful tool in many disciplines, especially in optomechanical applications and in the future quantum technologies.

\appendix
\setcounter{figure}{0} \global\long\def\thefigure{S\arabic{figure}}
\section{investigation method: Debye Integral and Lorentz-Mie Theory}
\label{appendix:A}

The investigation method is formulated concisely in this part. The
strongly focused incident laser beam is modeled by the generalized
vector Debye integral, and the generalized Lorentz-Mie theory (GLMT)
is used to solve the scattering problem of light field. Finally, the
time-averaged Maxwell stress tensor is used to calculate the force
exerted on the sphere particle.

The strongly focused beam is affected dramatically by the lens with
high numerical aperture (NA), and cannot be described by the expression
of paraxial beams. We model various doughnut beams, including linearly
polarized Laguerre-Gaussian (lpLG) beams, right circularly polarized
Laguerre-Gaussian (rcLG) beams and azimuthally polarized Gaussian
(apG) beam here, by the highly accurate generalized vector Debye integral
theory \cite{ChenPRE2009,Novotny2012principles}. The incident field
near the focus is formulated as:
\begin{eqnarray}
\mathbf{E}(\rho,\varphi,z) & = & -\frac{ikfe^{-ikf}}{2\pi}\int_{0}^{\theta_{\max}}d\theta\text{sin\ensuremath{\theta}}\int_{0}^{2\pi}d\phi[\mathbf{E}_{\infty}(\theta,\phi)\nonumber \\
 &  & \times e^{ik\rho\text{sin\ensuremath{\theta}}\cos(\phi-\varphi)+ikz\text{cos\ensuremath{\theta}}}],\label{eq:FocusedField}
\end{eqnarray}
where $\mathbf{E}_{\infty}(\theta,\phi)$ is the electric field vector
on the Gaussian reference sphere, and the exponential factor is the
phase accumulated during the light propagation. This expression has
been used to investigate the properties of strongly focused beams
and shows good accuracy \cite{NgPRL2010,Novotny2012principles}.

The scattering of incident beam is calculated by the GLMT \cite{Nieminen2007Optical,ChenPRE2009} and is formulated concisely here. In this theory, the incoming and outgoing fields are expanded in vector spherical
wavefunctions (VSWFs): 
\begin{eqnarray}
\mathbf{E}^{\mathrm{in}} & = & \stackrel[n=1]{\infty}{\sum}\stackrel[m=-n]{n}{\sum}[a_{mn}\mathbf{M}_{mn}(k\mathbf{r})+b_{mn}\mathbf{N}_{mn}(k\mathbf{r}),\\
\mathbf{E}^{\mathrm{out}} & = & \stackrel[n=1]{\infty}{\sum}\stackrel[m=-n]{n}{\sum}[p_{mn}\mathbf{M}_{mn}(k\mathbf{r})+q_{mn}\mathbf{N}_{mn}(k\mathbf{r}).
\end{eqnarray}
The coefficients of incoming and outgoing fields are related by the
T-matrix \textbf{$\mathbf{\mathbb{T}}$} of the particle as: 
\begin{equation}
\left(\begin{array}{c}
\mathbf{p}\\
\mathbf{q}
\end{array}\right)=\mathbf{\mathbb{T}}\left(\begin{array}{c}
\mathbf{a}\\
\mathbf{b}
\end{array}\right).
\end{equation}
For sphere or multi-layer sphere, $\mathbf{\mathbb{T}}$ is diagonal
and can be calculated directly or by iteration \cite{WuRS1991Electromagnetic,WuAO1997Improved}.

The force and torque exerted on the sphere particle are calculated
from time-averaged Maxwell stress tensor: 
\begin{eqnarray}
\bar{\mathcal{T}} & = & \frac{1}{2}\text{Re}[\varepsilon\mathbf{E^{*}E}+\frac{\mathbf{B^{*}B}}{\mu}-\frac{1}{2}\overleftrightarrow{1}(\varepsilon\mathbf{E}^{*}\cdot\mathbf{E}+\frac{\mathbf{B}^{*}\cdot\mathbf{B}}{\mu}),\\
\mathbf{F} & = & \varoint d\mathbf{S}\cdot\bar{\mathcal{T}},\\
\mathbf{\Gamma} & = & \varoint d\mathbf{S}\cdot(\mathbf{r}\times\bar{\mathcal{T}}).
\end{eqnarray}
They are then formulated by the incident and scattered field coefficients,
and finally we get, e.g,, the axial force and torque: 
\begin{eqnarray}
F_{z} & = & \frac{n_{m}P}{c}\frac{2}{P_{c}}\stackrel[n=1]{\infty}{\sum}\stackrel[m=-n]{n}{\sum}\{\frac{m}{n(n+1)}{\rm Re}(a_{mn}^{*}b_{mn}-p_{mn}^{*}q_{mn})\nonumber \\
 &  & -\frac{1}{n+1}[\frac{n(n+2)(n-m+1)(n+m+1)}{(2n+1)(2n+3)}]^{1/2}\nonumber \\
 &  & \times{\rm Im}(a_{mn}a_{mn+1}^{*}+b_{mn}b_{mn+1}^{*}-p_{mn}p_{mn+1}^{*}\nonumber \\
 &  & -q_{mn}q_{mn+1}^{*})\},\\
\tau_{z} & = & \frac{P}{kc}\frac{2}{P_{c}}\stackrel[n=1]{\infty}{\sum}\stackrel[m=-n]{n}{\sum}[m\nonumber \\
 &  & \times(a_{mn}a_{mn}^{*}+b_{mn}b_{mn}^{*}-p_{mn}p_{mn}^{*}-q_{mn}q_{mn}^{*})].
\end{eqnarray}
Among the last two equations, 
\begin{equation}
P_{c}=\stackrel[n=1]{\infty}{\sum}\stackrel[m=-n]{n}{\sum}(a_{mn}a_{mn}^{*}+b_{mn}b_{mn}^{*})
\end{equation}
is proportional to the incident power, $n_{m}$ is the refractive
index of the medium surrounding the particle and $P$ is the power
of the incident laser beam. Thus from the extinction and scattering
coefficients, we get the absorption coefficient of the particle: 
\begin{eqnarray}
c_{\mathrm{ext}} & = & -\frac{1}{P_{c}}\stackrel[n=1]{\infty}{\sum}\stackrel[m=-n]{n}{\sum}Re(a_{mn}p_{mn}^{*}+b_{mn}q_{mn}^{*}),\\
c_{\mathrm{sca}} & = & \frac{1}{P_{c}}\stackrel[n=1]{\infty}{\sum}\stackrel[m=-n]{n}{\sum}(p_{mn}p_{mn}^{*}+q_{mn}q_{mn}^{*}),\\
c_{\mathrm{abs}} & = & c_{\mathrm{ext}}-c_{\mathrm{sca}}.
\end{eqnarray}

\section{Phase diagram and Trapping stability in various doughnut beams}
\label{sec:Trap-stability}
\begin{figure*}
\includegraphics[width=11cm]{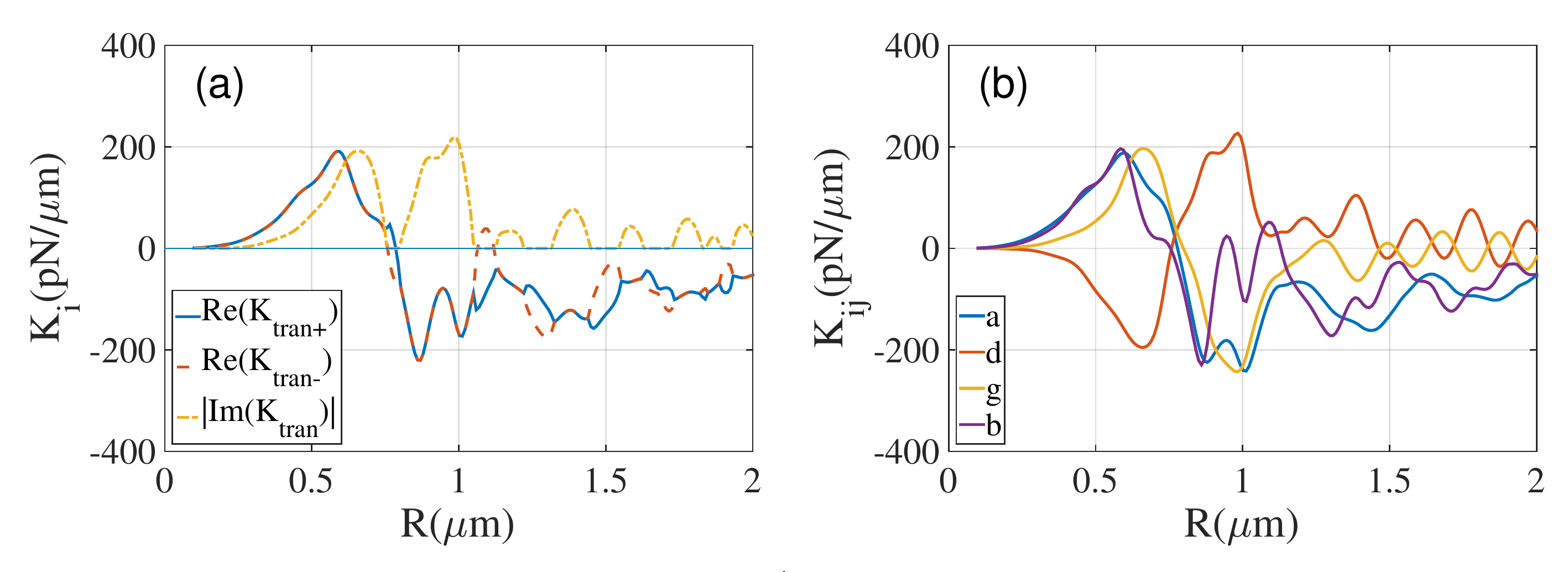}

\caption{(a) The transverse eigen force constant $K_{i}$ and (b) force constant
element $K_{ij}$ for a core-shell sphere trapped in a single ${\rm LG}_{03}$
beam polarized along $\boldsymbol{x}$. All beams are with power $P=100\ {\rm mW}$,
$\lambda=1064\ {\rm nm}$ and focused by lens with ${\rm NA}=0.95$.
All spheres are with core radius $r=20\ {\rm nm}$.}
\label{fig:K_lpLG03} 
\end{figure*}

\begin{figure*}
\includegraphics[width=15cm]{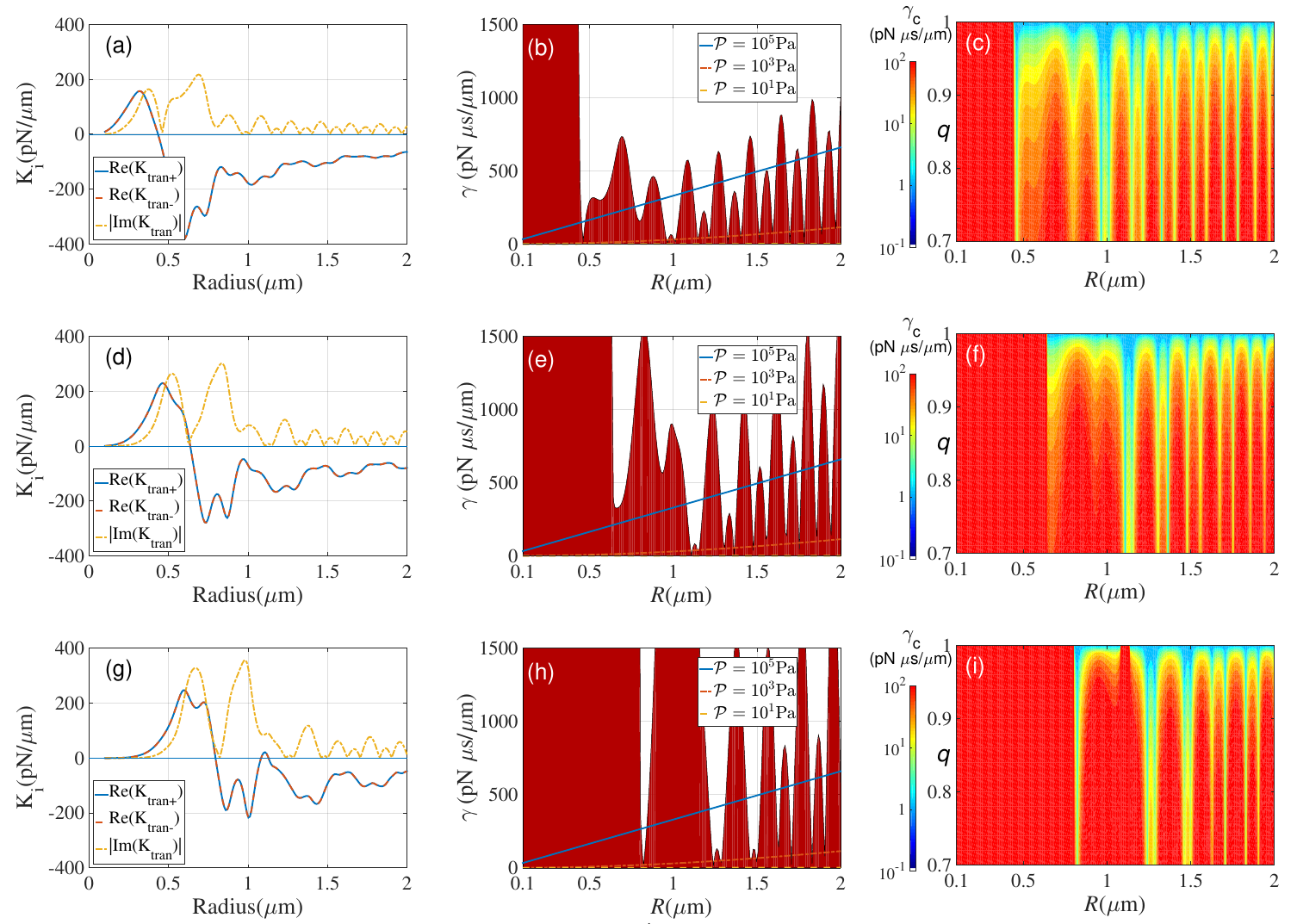} 

\caption{(a) The transverse eigen force constant $K_{i}$ and (b) the phase
diagram of a core-shell sphere in a single right circularly polarized
${\rm LG}_{01}$ beam. (c) Minimum friction $\gamma_{c}$ needed for
stable trapping in two counter-propagating ${\rm rcLG}_{01}$ beams
with different power ratio $q$. (d-f) The same as the first row,
for ${\rm rcLG}_{02}$ beams. (g-i) The same as the first row, for
${\rm rcLG}_{03}$ beams. All beams are with power $P=100\ {\rm mW}$,
$\lambda=1064\ {\rm nm}$ and focused by lens with ${\rm NA}=0.95$.
All spheres are with core radius $r=20\ {\rm nm}$.}
\label{fig:PhaseDiagramrcLG} 
\end{figure*}

Systematical investigation of the trapping stability in different
doughnut beams has been made and is shown in this part. For a sphere
in an optical trap, the force constant is a tensor $\mathbf{\mathbb{K}}$
with elements \cite{NgPRL2010} 
\begin{equation}
K_{ij}=\frac{\partial F_{i}}{\partial x_{j}},
\end{equation}
where $i,\ j=x,\ y,\ z$ are the Cartesian coordinates. The eigen
vectors of $\mathbf{\mathbb{K}}$ are the eigen-modes of sphere motion
and can be used to analyze its stability. For a sphere in a LG beam,
the force matrix can be written as 
\begin{equation}
\mathbf{\mathbb{K}}=\left[\begin{array}{ccc}
a & d & 0\\
g & b & 0\\
0 & 0 & c
\end{array}\right].
\end{equation}
The eigen motion modes along $z$ axis is independent of those on
the transverse $xy$ plane. Using the theory of the trapping stability
by optical vortex beam (i.e., LG beam) \cite{NgPRL2010}, various
LG beams with different $l$ and polarizations have been investigated
here.

Especially, the case of a single $\boldsymbol{x}$ polarized ${\rm LG}_{03}$
beam are shown in Fig.~\ref{fig:K_lpLG03}(a) for eigen force constant,
in Fig.~\ref{fig:K_lpLG03}(b) for force constant elements and in Fig.~2(c)
of the main text for phase diagram. It is noticed that, different
from the phase diagram of ${\rm LG}_{01}$ and ${\rm LG}_{02}$ beam,
there is a region which is always unstable despite of the environment
damping at $R\sim1.1{\rm \mu m}$ {[}see Fig.~2(c){]}. It can be seen
from Fig.~\ref{fig:K_lpLG03}(b) that the force constant $b=K_{yy}$
is positive at $R\sim1.1\ {\rm \mu m}$.

The rcLG beams are different from lpLG beam as they have no stable
window when $\gamma=0$, as shown in Fig.~\ref{fig:PhaseDiagramrcLG}(b)(e)(h)
for $l=1$, 2, 3, respectively. For different $q$, the minimum $\gamma_{c}$
needed to keep stable is shown in Fig.~\ref{fig:PhaseDiagramrcLG}(c)(f)(i).

\section{Approximate expression of the intensity near the focus}

The field of the strongly focused beam near the focus determines the
absorption coefficients of the levitated core-shell sphere. Although
there is no analytic expression of the field of the strongly focused
beam, an approximate analytical expression near the beam focus will
be helpful both for theory and experimental investigation. Our analytical
approximations are shown in this part.

The field distribution of focused LG beam is expressed with vector
Debye integral theory with Eq.~(\ref{eq:FocusedField}). When $z=0$,
and 
\begin{equation}
\delta\triangleq k\rho\ll1,
\end{equation}
we expand the field near the focus with Taylor series. For an $\hat{\boldsymbol{x}}$
direction linearly polarized incident ${\rm LG}_{0l}$ beam, the light
intensity $I_{l}=\frac{1}{2}\varepsilon_{m}\varepsilon_{0}v_{m}|\mathbf{E}_{l}|^{2}$
near the focus can be expressed as $I_{l}=A_{l}\delta^{n_{l}}$, $n_{l}=2(l-2)$.
The absorption coefficients of a small sphere with radius $r$ can
be approximated as

\begin{equation}
c_{\mathrm{abs}}=\int_{V}\eta_{l}\alpha_{l}I_{l}dV/P_{\mathrm{inc}},
\end{equation}
where $\alpha_{l}$ is the light attenuation coefficient in sphere,
$I_{l}$ is the light intensity and $V$ is domain of the sphere core
with radius $r$. $\eta_{l}$ is a correction factor considering the
affection such as the changing of light field after scattering. The
suppression coefficients ratio are defined as $\xi=c_{\mathrm{abs}}/c_{\mathrm{abs},0}$
and follow different power-law as $\xi_{l}=\xi_{l0}\delta^{n_{l}}$,
$n_{l}=2(l-2)$. Choosing $\eta_{0}=1.0$, $\eta_{1}=1.3$, $\eta_{2}=0.9$,
$\eta_{3}=0.45$, and $\eta_{4}=0.40$ when ${\rm NA}=1.0$, we get
$\xi_{10}=0.21$, $\xi_{20}=0.034$, $\xi_{30}=0.0032$, and $\xi_{40}=1.8\times10^{-4}$.
The approximate analytical results of $\xi_{l}$ are plotted in Fig.~3(a)
with blue lines and match the numerical results well.

Absorption suppression for rcLG beams shows similar behavior as those
in lpLG beams. The absorption scale with $r$ can also be analyzed
as before. The suppression coefficients follow different power-law
as $\xi_{l}=\xi_{l0}\delta^{n_{l}}$, $n_{l}=2l$. The approximate
results are also plotted in Fig.~\ref{fig:FreqQ}(a). Though the suppression
is stronger in rcLG beams than that in lpLG beams, the minimum $\gamma_{c}$
needed limits quality factor of the mechanical oscillation. So, the
linear polarized LG beams are better for optomechanical application.

The apG beam also have zeros intensity in the beam focus. The phase
diagram for a sphere in apG beam is shown in Fig.~2(b). Above the transition
radius $R_{{\rm tran}}=0.33\ {\rm \mu m}$, the sphere is always stable.
For apG beam, the suppression coefficient ratio can be written as
$\xi_{a}=\xi_{a0}\delta^{2}$. It is also plotted in Fig.~3(a).

\section{quality factor $Q$ in various doughnut beams and some discussions\label{sec:Light-absorbtion-reducement}}

\begin{figure*}
\includegraphics[width=15cm]{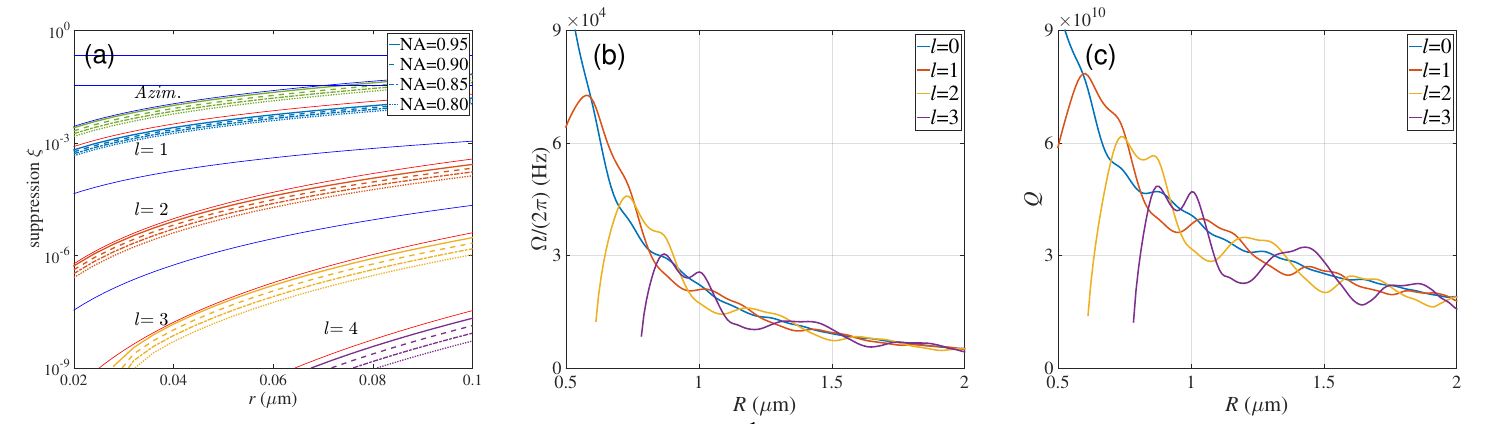}

\caption{(a) Light absorption suppression ratio $\xi$ for a core-shell sphere
at the center of azimuthally and right circularly polarized beams
($\mathrm{LG}_{0l}$ beams) relative to linearly polarized Gaussian
beams, with different core radius $r$ and numerical aperture $\mathrm{NA}$.
Red solid lines are for our analytical results. Blue lines are for
linearly polarized LG beams. Other parameters are the same as Fig.~3(a).
(b) The optomechanical frequency $\Omega/(2\pi)$ and (c) mechanical
quality factor $Q$ of a sphere in different lpLG beams along $\boldsymbol{x}$
with power $100\ {\rm mW}$ under pressure $10^{-6}\ {\rm Pa}$.}
\label{fig:FreqQ} 
\end{figure*}

Doughnut beams have different field distributions ( especially, different
radius of their bright rings) and thus have different trapping potential.
The mechanical oscillation frequency $\Omega$ and quality factor
$Q$ are investigated systematically. generally, larger radius of
bright rings lead to smaller $\Omega$ and $Q$. The case for a spheres
in linearly polarized ${\rm LG}_{0l}$ beams with different $l$ is
shown in Fig.~\ref{fig:FreqQ}.

\begin{acknowledgments}
We are grateful to Prof. Zhifang Lin for sharing the numerical code.
We thank Prof. Jack Ng for the helpful discussion. This work was supported by the NSFC grants (No. 11374032). J.C. is also supported by Shanxi Science and Technology Department through 2014011008-1, 2014021004 and NSFC through 11404201, 11674204.
\end{acknowledgments}

\end{document}